\documentclass{article}
\usepackage{slashed,feynmf,epsfig,amsmath,amssymb,enumitem}
\usepackage[papersize={8.5in,11in}]{geometry}
\begin{document}
\title{Theory of Electroweak Gauge Interactions}
\author{J. W. Moffat\\~\\
Perimeter Institute for Theoretical Physics, Waterloo, Ontario N2L 2Y5, Canada\\
and\\
Department of Physics and Astronomy, University of Waterloo, Waterloo,\\
Ontario N2L 3G1, Canada}
\maketitle
\begin{abstract}
The conditions obtained by Salam for a general gauge theory to be renormalizable are derived. They require that in a gauge invariant formalism the bare boson mass associated with the massive non-Abelian vector field is zero. A solution to the origin of masses in electroweak theory is proposed in which the physical masses of the $W$ and $Z$ bosons and the quarks and leptons are derived from their self-energies according to a self-consistent scheme.
\end{abstract}

\begin{fmffile}{ewunifigs}

\section{Introduction}

 The Higgs particle has not yet been conclusively detected. If it is not detected at the LHC accelerator, then we must consider revising at a fundamental level the electroweak (EW) model of Weinberg and Salam~\cite{Weinberg,Salam,Aitchison}. This may require a revision of our ideas about QFT. A previously published EW theory without a Higgs particle and a quantum gravity theory~\cite{Moffat2,Moffat3,Moffat4,Moffat5} were based on nonlocal interactions. The EW theory led to finite amplitudes and cross sections that can be tested at the LHC.

In a recent paper~\cite{Moffat6,Moffat7}, an effective EW model was based on a gauge invariant action with {\it local interactions} using a Stueckelberg formalism~\cite{Stueckelberg,Ruegg}. The gauge invariance of the Lagrangian led to a renormalizable EW theory, provided the scalar fields in the Lagrangian decouple at high energies rendering a scalar spin-0 boson undetectable at present accelerator energies. This decoupling could only be maintained in a restricted energy range; the theory was only an effective theory and was not ultraviolet (UV) complete.

In the following, we will explore further the breaking of EW symmetry by attempting to derive the physical masses of particles through a self-consistent scheme within a renormalizable EW theory. We begin by reviewing the conditions that have to be met for a general gauge theory to be renormalizable. This leads to the general conditions derived by Salam~\cite{Salam2} for a gauge theory with massive vector bosons to be renormalizable. We then pursue the idea that the elementary quarks, leptons and the $W$ and $Z$ bosons acquire their masses from the basic self-energies of the primary fields. At the tree graph level, the bare particles are massless. Then as the particles interact their quantum field self-energies generate masses of the dressed particles. Following Nambu-Jona-Lasinio~\cite{Nambu}, the masses of the elementary particles arise from the proper self-energies using a Hartree-Fock self-consistent approximation in which the boson and fermion bare masses are zero. The theory contains only the currently observed particles, namely, twelve quarks and leptons, the charged $W$ boson, the neutral $Z$ boson and the massless photon and gluons without the Higgs particle. In contrast to the standard EW theory, in which the masses of the elementary particles are obtained already at the tree graph level by introducing a physical scalar degree of freedom and postulating a spontaneous symmetry breaking of the classical vacuum, the present solution to the problem of the origin of particle masses resides in the quantum interactions of the particle fields without the prediction of a physical scalar Higgs particle in the particle spectrum. The tree graph, Born approximation calculation of the longitudinally polarized $W_LW_L\rightarrow W_LW_L$ scattering amplitude is shown to not violate the unitarity bound within the renormalized self-consistent, fully dressed particle approximation scheme.

\section{Gauge Theories and Renormalizability}

Let us review the structure of gauge theories and their renormalizability. We will use the metric convention, $\eta_{\mu\nu}={\rm diag}(+1,-1,-1,-1)$ and set $\hbar=c=1$. In standard perturbation theory, we solve by successive approximations starting with the bare mass $M_0$ and bare coupling constant $g_0$. The states in the S-matrix correspond to ``undressed'' bare states. However, we can entertain the idea that there are physical non-trivial solutions which cannot be thus obtained. In fact, there exist solutions with physical mass $M\not=0$ when the bare mass, $M_0=0$. We can understand this by considering a self-consistent Hartree-Fock type of procedure~\cite{Nambu}. In standard perturbation theory, we compose the Lagrangian from the free Lagrangian ${\cal L}_0$ and the interaction Lagrangian ${\cal L}_I$:
\begin{equation}
{\cal L}={\cal L}_0+{\cal L}_I.
\end{equation}
Instead of diagonalizing ${\cal L}_0$ and treating the interaction part as a perturbation, we introduce the self-energy Lagrangian ${\cal L}_{\rm self}$ and split ${\cal L}$ as
\begin{equation}
\label{ShiftedLagrangian}
{\cal L}=({\cal L}_0+{\cal L}_{\rm self})+({\cal L}_I-{\cal L}_{\rm self})={\cal L}_0'+{\cal L}'_I.
\end{equation}
We can now define a new vacuum and a complete set of ``quasi-particle'' states for which each particle is an eigenmode of ${\cal L}_0'$. We now solve ${\cal L}_{\rm self}$ as a perturbation and determine ${\cal L}_{\rm self}$ without producing additional self-energy effects. The self-consistent nature of the procedure allows the self-energy to be calculated by perturbation theory with the fields defined by a new vacuum which are already subject to the self-energy interaction.

Let us begin by considering the Lagrangian in what we shall call the ``bare'' particle representation in which ${\cal L}={\cal L}(M_0,g_0)$ where $M_0$ and $g_0$ are the bare mass and charge. The Lagrangian is given by
\begin{equation}
\label{GaugeLagrangian}
{\cal L}={\cal L}_f+{\cal L}_B,
\end{equation}
where
\begin{equation}
\label{FermionLagrangian}
{\cal L}_f=\sum_\psi {\bar\psi}i\slashed{D}\psi-m_f{\bar\psi}\psi,
\end{equation}
and
\begin{equation}
{\cal L}_B=-\frac{1}{4}{\rm Tr}[F^{\mu\nu}F_{\mu\nu}].
\end{equation}
Here, we define
\begin{equation}
F_{\mu\nu}=D_\nu V_\mu-D_\mu V_\nu,
\end{equation}
where
\begin{equation}
D_\mu=\partial_\mu+ig_0V_\mu
\end{equation}
is the covariant derivative. Moreover, we have
\begin{equation}
V_\mu=T^aV^a_\mu,
\end{equation}
where $T^a$ is a set of non-Abelian generators which obey
\begin{equation}
[T^a,T^b]={C_c}^{ab}T_c.
\end{equation}

The fermion Lagrangian (\ref{FermionLagrangian}) is invariant under the transformation
\begin{equation}
\psi'=U\psi,
\end{equation}
where
\begin{equation}
U=\exp(ig_0T^ab^a),
\end{equation}
and the $b^a$ are fields to be specified later.
The vector field $V_\mu$ transform as
\begin{equation}
V_\mu'=UV_\mu U^{-1}+\frac{i}{g_0}(\partial_\mu U)U^{-1}.
\end{equation}
This assures that
\begin{equation}
{\bar\psi}'D'_\mu\psi'={\bar\psi}D_\mu\psi
\end{equation}
is satisfied. We also have that
\begin{equation}
F'_{\mu\nu}=UF_{\mu\nu}U^{-1}.
\end{equation}

We can add to the Lagrangian (\ref{GaugeLagrangian}) the non-gauge invariant mass term
\begin{equation}
\label{Massterm}
{\cal L}_M=-\frac{1}{2}M_0^2{\rm Tr}[V_\mu V^\mu],
\end{equation}
where $M_0$ is the bare mass of the vector boson field $V_\mu$. For the weak interactions we will also have to account for non-gauge invariant chiral fermion mass terms.

To study the renormalizability of our theory, we introduce the Stueckelberg fields~\cite{Stueckelberg,Ruegg}:
\begin{equation}
V_\mu=G_\mu-\frac{1}{M_0}\partial_\mu\sigma,
\end{equation}
where
\begin{equation}
G_\mu=G^a_\mu T^a,\quad \sigma=\sigma^aT^a.
\end{equation}
Now the transformation matrix $U$ has the form
\begin{equation}
\label{Umatrix}
U=\exp(ig_0\sigma/M_0).
\end{equation}
We now change $\psi$ to $\psi'$ and $V^\mu$ to $V^{'\mu}$ using the gauge transformations. Under these transformations
\begin{equation}
{\cal L}(\psi',V_\mu')={\cal L}(\psi,V_\mu),
\end{equation}
and we maintain gauge invariance, whereas (\ref{Massterm}) is not gauge invariant.

We have now in the weak coupling limit $g\rightarrow 0$ where asymptotic states are defined:
\begin{equation}
V_\mu'=G_\mu+O(g)
\end{equation}
and
\begin{equation}
V_\mu^{'\rm in}=G_\mu^{\rm in},\quad \psi^{'\rm in}=\psi^{\rm in}.
\end{equation}
The S-matrix formed from the new variables has contributions from two parts of the Lagrangian. The part
\begin{equation}
{\cal L}(G_\mu^{\rm in},\psi^{\rm in}),
\end{equation}
yields only renormalizable infinities, because the derivative couplings of $\sigma$ have been eliminated, while those that are generated by the mass term (\ref{Massterm}):
\begin{equation}
{\cal L}(G_\mu^{\rm in},\sigma^{\rm in})
\end{equation}
generate bad exponential infinities due to the occurrence of the transformation matrix (\ref{Umatrix})~\cite{Salam2}.

The non-renormalizable exponential infinities can be avoided if either
\begin{equation}
\label{Salamcondition}
M_0=0,
\end{equation}
or
\begin{equation}
\label{Condition1}
{\rm Tr}\biggl[\frac{M_0^2}{g_0^2}\partial^\mu U\partial_\mu U^{-1}-(\partial_\mu\sigma^{\rm in})^2\biggr]=0,
\end{equation}
and
\begin{equation}
\label{Condition2}
{\rm Tr}\biggl[G_\mu^{\rm in}(U^{-1}\partial^\mu U-ig_0\partial^\mu\sigma^{\rm in})\biggr]=0.
\end{equation}
This powerful theorem, derived by Salam~\cite{Salam2}, determines the conditions required to obtain a renormalizable boson gauge theory. The problem is to find the gauge transformations that satisfy (\ref{Condition1}) and (\ref{Condition2}). If the condition (\ref{Salamcondition}) is satisfied then the gauge bosons in the bare particle representation are massless and the theory is renormalizable. On the other hand, if conditions (\ref{Condition1}) and (\ref{Condition2}) are satisfied, then a non-Abelian gauge theory is renormalizable.

For a massive neutral vector boson interacting with quarks and leptons there is only one $\sigma$ field, $U=\exp(ig_0\sigma/M_0)$ is Abelian and both (\ref{Condition1}) and (\ref{Condition2}) are satisfied and the theory is renormalizable. In general, the conditions (\ref{Condition1}) and (\ref{Condition2}) can be satisfied if
\begin{equation}
Tr[T^aT^b]\equiv 0.
\end{equation}
However, in general we have for non-Abelian gauge theory
\begin{equation}
{\rm Tr}[T^aT^b]=\lambda\delta^{ab},
\end{equation}
so that conditions (\ref{Condition1}) and (\ref{Condition2}) cannot be satisfied and the theory is not renormalizable.

The above arguments involved the vector boson interactions and the boson mass terms. Clearly any other non-gauge invariant interaction term involving the fermion mass terms and ${\cal L}_f(\psi)$ will get transformed into ${\cal L}_f(U\psi)$ and the exponential contributions involving the
$\sigma$ fields will produce non-renormalizable infinities.

\section{The Gauge Invariant Electroweak Lagrangian}

We now consider a theory based on the local $SU_L(2)\times U_Y(1)$ invariant Lagrangian that includes leptons and quarks (with the color degree of freedom of the strong interaction group $SU_c(3)$) and the boson vector fields that arise from gauging the $SU_L(2)\times U_Y(1)$ symmetry:
\begin{equation}
\label{Lag}
{\cal L}_\mathrm{EW}={\cal L}_f+{\cal L}_W+{\cal L}_B+{\cal L}_I.
\end{equation}
${\cal L}_f$ is the free fermion Lagrangian consisting of massless kinetic terms for each fermion:
\begin{equation}
{\cal L}_f=\sum_\psi\bar\psi i\slashed\partial\psi=\sum_{q^L}\bar q^Li\slashed\partial q^L+\sum_f\bar\psi^Ri\slashed\partial
\psi^R,
\end{equation}
where the fermion fields have been rewritten as $SU_L(2)$ doublets:
\begin{equation}
q^L\in\left[\begin{pmatrix}\nu^L\\e^L\end{pmatrix},\begin{pmatrix}u^L\\d^L\end{pmatrix}_{r,g,b}\right]
\end{equation}
and U(1)$_Y$ singlets, and we have suppressed the fermion generation indices. We have written $\psi_{L,R}=P_{L,R}\psi$ where $P_{L,R}=\frac{1}{2}(1\mp\gamma_5)$. The Abelian kinetic contribution is given by
\begin{equation}
{\cal L}_B=-\frac{1}{4}B^{\mu\nu}B_{\mu\nu},
\end{equation}
where
\begin{equation}
B_{\mu\nu}=\partial_\mu B_\nu-\partial_\nu B_\mu.
\end{equation}
The non-Abelian contribution is
\begin{equation}
{\cal L}_W=-\frac{1}{4}W_{\mu\nu}^aW^{a\mu\nu},
\end{equation}
where
\begin{equation}
W^a_{\mu\nu}=\partial_\mu W_\nu^a-\partial_\nu W_\mu^a+g_0f^{abc}W_\mu^bW_\nu^c.
\end{equation}

The $SU(2)$ generators satisfy the commutation relations
\begin{equation}
[T^a,T^b]=if^{abc}T^c,~~~~~\mathrm{with}~~~~~T^a=\frac{1}{2}\sigma^a.
\end{equation}
Here, $\sigma^a$ are the Pauli spin matrices and $f^{abc}=\epsilon^{abc}$. The fermion--gauge boson interaction terms are contained in
\begin{equation}
{\cal L}_I=-g_0J^{a\mu}W_\mu^a-g_0'J_Y^\mu B_\mu,
\end{equation}
where the $SU(2)$ and hypercharge currents are given by
\begin{equation}
J^{a\mu}=\sum_{q^L}\bar{q}^L\gamma^\mu T^aq^L,~~~~~\mathrm{and}~~~~~J_Y^\mu=\sum_\psi\frac{1}{2}Y\bar\psi\gamma^\mu\psi,
\end{equation}
respectively. The last sum is over all left and right-handed fermion states with hypercharge factors $Y=2(Q-T^3)$. We also define for notational convenience, $\slashed W=\gamma^\mu W_\mu^aT^a.$

The Lagrangian (\ref{Lag}) is invariant under the following local gauge transformations (to order $g,g'$):
\begin{align}
\delta W_\mu^a=\partial_\mu\theta^a+g_0f^{abc}\theta^b W_\mu^c,~~~&~~~\delta B_\mu=\partial_\mu\beta,\nonumber\\
\delta\psi^L=-\left(ig_0T^a\theta^a+ig_0'\frac{Y}{2}\beta\right)\psi^L,~~~&~~~\delta\psi^R=-ig_0'\frac{Y}{2}
\beta\psi^R,
\label{localtransformations}
\end{align}
giving us an $SU_L(2)\times U_Y(1)$ invariant Lagrangian. It can also be shown to be invariant under BRST transformations generated by replacing the infinitesimal fields $\theta^a$ and $\beta$ by
\begin{equation}
\theta^a\rightarrow-c^a\lambda\xi,~~~~~\beta\rightarrow-\eta\lambda_0\xi,
\end{equation}
where $\lambda$ and $\lambda_0$ are infinitesimal Grassmann constants~\cite{Becchi,Tyutin}.

Quantization is accomplished via the path integral formalism, which gives the expectation value of operators ${\cal O}[\phi]$ as a sum over all field configurations weighted by the exponential of the classical action:
\begin{equation}
\label{expectationvalue}
\left<T({\cal O}[\phi])\right>\propto\int[D\bar\psi][D\psi][DW][DB]\mu_\mathrm{inv}[\bar\psi,\psi,B,W]{\cal O}[\phi]
\exp\left(i\int d^4x{\cal L}_\mathrm{EW}\right),
\end{equation}
where $\mu_\mathrm{inv}$ is the invariant measure.  We introduce gauge fixing terms:
\begin{equation}
\label{Gaugefix}
{\cal L}_\mathrm{gf}=-\frac{1}{2\xi}(\partial_\mu B^\mu)^2-\frac{1}{2\xi}(\partial_\mu W^{a\mu})^2.
\end{equation}
Here, the choice of the gauge parameter $\xi$ could be different for each gauge field, but for simplicity we have chosen the same gauge condition for the Abelian and non-Abelian gauge bosons. As we require gauge invariant results, this constraint should not cause any physical prediction to pick up a dependence on the gauge parameter $\xi$. We ensure this by introducing auxiliary ghost fields $\eta$ and $c$ into the theory.

We have the final form of the path integral in the quantized theory:
\begin{equation}
\left<T({\cal O}[\phi])\right>\propto\int[D\bar\psi][D\psi][DW][DB][D\bar\eta][D\eta][D\bar c][Dc]{\cal O}[\phi]\exp(iS_\mathrm{eff}),
\end{equation}
where the effective action is given by
\begin{equation}
\label{effaction}
S_\mathrm{eff}=\int d^4x({\cal L}_\mathrm{EW}+{\cal L}_\mathrm{gf}+{\cal L}_\mathrm{ghost})=\int d^4x({\cal L}_F+{\cal L}_I).
\end{equation}
Here, we have separated the Lagrangian into a quadratic part ${\cal L}_F$ and an interaction part ${\cal L}_I$.

Finally, we look at diagonalizing the charged sector and mixing in the neutral boson sector. This will be performed in terms of the observed physical coupling constants $g$ and $g'$ in the dressed particle basis. If we write
\begin{equation}
W^\pm=\frac{1}{\sqrt{2}}(W^1\mp iW^2)
\end{equation}
as the physical charged vector boson fields, then we get the fermion interaction terms:
\begin{equation}
-\frac{g}{\sqrt{2}}(J_\mu^+W^{+\mu}+J_\mu^-W^{-\mu}),
\end{equation}
where the charged current is given by
\begin{equation}
J_\mu^\pm=J_{1\mu}^\pm\pm iJ_{2\mu}^\pm=\sum_{q_L}\bar q^L\gamma_\mu T^\pm q^L~~~~~{\rm implying}~~~~~J_\mu^+=\sum_{q_L}(\bar\nu^L\gamma_\mu e^L+\bar u^L\gamma_\mu d^L).
\end{equation}
In the neutral sector, we can mix the fields in the usual way:
\begin{equation}
Z_\mu=\cos\theta_wW_\mu^3-\sin\theta_wB_\mu~~~\mathrm{and}~~~A_\mu=\cos\theta_wB_\mu+\sin\theta_wW_\mu^3,
\label{eq:2.35}
\end{equation}
where $\theta_w$ denotes the weak mixing angle.

We define the usual trigonometric relations
\begin{equation}
\label{Trigrelations}
\sin^2\theta_w=\frac{g'^2}{g^2+g'^2}~~~\mathrm{and}~~~\cos^2\theta_w=\frac{g^2}{g^2+g'^2}.
\end{equation}
The neutral current fermion interaction terms now look like:
\begin{equation}
-gJ^{3\mu}W_\mu^3-g'J_Y^\mu B_\mu=-(g\sin\theta_wJ^{3\mu}+g'\cos\theta_wJ_Y^\mu)A_\mu-(g\cos\theta_wJ^{3\mu}-g'\sin\theta_wJ_Y^\mu)Z_\mu.
\end{equation}
If we identify the resulting $A_\mu$ field with the photon, then we have the unification condition:
\begin{equation}
e=g\sin\theta_w=g'\cos\theta_w
\end{equation}
and the electromagnetic current is
\begin{equation}
J_\mathrm{em}^\mu=J^{3\mu}+J_Y^\mu,
\end{equation}
where $e$ is the charge of the proton. Note that the coupling now looks like $(Q-T^3)+T^3=Q$ and we only get coupling of the photon to charged fermions at tree level. We can then identify the neutral current:
\begin{equation}
J_\mathrm{NC}^\mu=J^{3\mu}-\sin^2\theta_wJ_\mathrm{em}^\mu,
\end{equation}
and write the fermion-boson interaction terms as
\begin{equation}
L_I=-\frac{g}{\sqrt{2}}(J_\mu^+W^{+\mu}+J_\mu^-W^{-\mu})-g\sin\theta_wJ_\mathrm{em}^\mu A_\mu-\frac{g}{\cos\theta_w}J_\mathrm{NC}^\mu Z_\mu.
\end{equation}
This, along with the suitably rewritten boson interaction terms, gives the usual vertices of the local point theory.

We have not included in the Lagrangian $(\ref{Lag})$ the standard scalar field Higgs contribution:
\begin{equation}
\label{scalarLagrangian}
{\cal L}_\phi=\large\vert(i\partial_\mu - gT^aW^a_\mu - g'\frac{Y}{2}B_\mu)\phi\large\vert^2-V(\phi),
\end{equation}
where $\vert...\vert^2=(...)^\dagger(...)$. Moreover,
\begin{equation}
\label{potential}
V(\phi)=\mu_H^2\phi^\dagger\phi +\lambda_H(\phi^\dagger\phi)^2,
\end{equation}
and $\mu^2_H < 0$ and $\lambda_H > 0$.

\section{Feynman Rules And The Gauge Boson Self-Energy}

By employing the gauge fixing conditions (\ref{Gaugefix}), we obtain the vector boson propagator in the ``bare''particle representation with $M_{0V}=0$:
\begin{equation}
\label{barepropagator}
iD^V_{\mu\nu}(p^2)=i\frac{-\eta_{\mu\nu}+(1-\xi)\frac{p_\mu p_\nu}{p^2}}{p^2+i\epsilon}.
\end{equation}
The boson propagator in the physical ``dressed'' particle representation is given by
\begin{equation}
i\Delta^V_{\mu\nu}(p^2)=i\frac{-\eta_{\mu\nu} +\frac{(1-\xi)\frac{p_\mu p_\nu}{M_V^2}}{p^2-\xi M_V^2}}{p^2-M_V^2+i\epsilon},
\end{equation}
where $M^2_V$ is the boson mass identified with the the $Z$ and $W$ proper self-energies. For a finite value of the gauge parameter $\xi$, the propagators $D^V_{\mu\nu}$ and $\Delta^V_{\mu\nu}$ behave as $1/p^2$ for $p^2\rightarrow\infty$ and the theory is renormalizable. The fermion propagator in the dressed particle representation is given by
\begin{equation}
iS(p)=\frac{i}{{\slashed p}-m_f+i\epsilon}=\frac{i({\slashed p}+m_f)}{p^2-m_f^2+i\epsilon},
\end{equation}
where $m_f=\Sigma_f$ is the fermion mass identified with the fermion proper self-energy and the bare fermion mass $m_{0f}=0$. These identifications of the boson and fermion masses in the propagators are interpreted in a non-perturbative sense in terms of energy gap equations in our self-consistent scheme. The interpretation of these energy gap equations should be based on a development of our EW theory using a renormalization group flow formalism.

We will employ the propagator for the bosons in the bare particle representation to calculate the proper self-energies in the gauge invariant limit $M_{0V}=0$. However, in the calculation, we use the physical dressed fermion mass $m_f$. The vacuum polarization tensor $\Pi_{\mu\nu}$ obtained from the lowest order vector boson self-energy diagram is of the form:
\begin{equation}
i\Pi_{\mu\nu}(q^2)=i\Pi^T(q^2)\biggl(\eta_{\mu\nu}-\frac{q_\mu q_\nu}{q^2}\biggr)-i\Pi^L(q^2)\frac{q_\mu q_\nu}{q^2},
\end{equation}
where $\Pi^T$ and $\Pi^L$ denote the transverse and longitudinal coefficients of the vacuum polarization tensor. The modification of the lowest order boson propagator in the gauge $\xi=1$ is given by
\begin{equation}
\label{ModifiedProp}
\frac{-i\eta_{\mu\nu}}{q^2}\rightarrow \frac{-i\eta_{\mu\nu}}{q^2}+\biggl(\frac{-i}{q^2}\biggr)i\Pi_{\mu\nu}(q^2)\biggl(\frac{-i}{q^2}\biggr).
\end{equation}
By means of a regularization of the boson propagator~\cite{Evens1991}, we find the result
\begin{equation}
\label{Pitransverse}
\Pi(Q^2)\equiv\Pi^T(Q^2)=Q^2\frac{2\alpha_g}{\pi}\int^{1/2}_0dxx(1-x)E_1\biggl[\frac{1}{1-x}\frac{m_f^2}{\Lambda_B^2}+x\frac{Q^2}{\Lambda_B^2}\biggr],
\end{equation}
where $E_1$ is the exponential integral:
\begin{equation}
E_1(z)\equiv \int^\infty_z
dt\frac{\exp(-t)}{t}=-\ln(z)-\gamma-\sum_{n=1}^\infty\frac{(-z)^n}{nn!},
\end{equation}
and $\gamma$ is the Euler-Mascheroni number $\gamma=0.57726$. Moreover, $Q^2=-q^2$, $\alpha_g=g^2/4\pi$ and $\Lambda_B$ is the boson cutoff. Note that $\Pi(0)\equiv\Pi^T(0)=0$ in accordance with the underlying gauge invariance.

We obtain from (\ref{Pitransverse}) for small $Q^2$:
\begin{equation}
\label{smallQ}
\Pi(Q^2)\sim Q^2\biggl[\frac{\alpha_g}{3\pi}\ln\biggl(\frac{\Lambda_B^2}{m_f^2}\biggr)-\frac{\alpha_g}{15\pi}\frac{Q^2}{m_f^2}\biggr].
\end{equation}
An asymptotic expansion of the integral in (\ref{Pitransverse}) for large $Q^2$ and $\Lambda^2_B$ gives the result~\cite{Evens1991,Halzen}:
\begin{equation}
\label{Pi}
\Pi(Q^2)\sim Q^2\biggl\{\frac{\alpha_g}{3\pi}\ln\biggl(\frac{\Lambda_B^2}{m_f^2}\biggr)-\frac{2\alpha_g}{\pi}\int_0^1dxx(1-x)\ln\bigg[1+
\frac{Q^2x(1-x)}{m_f^2}\biggr]+O\biggl(\frac{\ln\biggl(\Lambda_B^2/m_f^2\biggr)}{\Lambda_B^2}\biggr)\biggr\}.
\end{equation}
We obtain from (\ref{Pi}) for large $Q^2=-q^2$:
\begin{equation}
\label{Piequation}
\Pi(Q^2)=Q^2\frac{\alpha_g}{3\pi}\ln\biggl(\frac{\Lambda_B^2}{Q^2}\biggr).
\end{equation}

Consider Rutherford scattering in quantum electrodynamics (QED). The amplitude for small $Q^2$ obtained from (\ref{ModifiedProp}) and (\ref{smallQ}) is given by
\begin{equation}
-i{\cal A}(Q^2)=(ie_0{\bar u}\gamma_0 u)\biggl(\frac{i}{Q^2}\biggr)\biggl[1-\frac{\alpha}{3\pi}\ln\biggl(\frac{\Lambda^2_\gamma}{m^2_e}\biggr)
+\frac{\alpha}{15\pi}\frac{Q^2}{m^2_e}+O(e_0^4)\biggr](-iZe_0),
\end{equation}
where $Ze_0$ is the static nuclear charge and $\alpha=e_0^2/4\pi$ is the fine structure constant. We can rewrite this result as
\begin{equation}
-i{\cal A}(Q^2)=(ie_R{\bar u}\gamma_0 u)\biggl[\frac{i}{Q^2}\biggl(1+\frac{e_R^2}{60\pi^2}\frac{Q^2}{m_e^2}\biggr)\biggr](-iZe_R),
\end{equation}
where
\begin{equation}
e_R=e_0\biggl[1-\frac{e_0^2}{12\pi^2}\ln\biggl(\frac{\Lambda_\gamma^2}{m_e^2}\biggr)\biggr]^{1/2}=e_0Z^{-1}_1Z_2Z_3^{1/2},
\end{equation}
is the renormalized charge and by the Ward-Takahashi identity: $Z_1=Z_2$. The renormalized charge $e_R$ leads to the well-known contribution to the Lamb shift:
\begin{equation}
\Delta E_{nl}=-\frac{8\alpha_R^3}{15\pi n^3}Ry\delta_{l0},
\end{equation}
where $\alpha_R=e_R^2/4\pi$ is the renormalized fine structure constant.

Whereas in our QFT renormalization is performed in the bare Lagrangian representation for which $M_{0V}=0$ and $m_{0f}=0$, we will develop our EW theory using the dressed particle representation with zero bare masses and finite boson and fermion proper self-energies. In the bare particle representation the results of the loop calculation to all orders in the bare coupling constant $g_0$ is
\begin{equation}
g=g_0[1+g_0^2C_1(Q^2)+g_0^4C_2(Q^2)+...]_{Q^2=\mu^2},
\end{equation}
where $q^2\equiv -Q^2=-\mu^2$ is a particular value of the virtual boson's momentum appropriate to an experiment, and $C_1(Q^2)$ is directly related to the infinite quantity $\Pi(Q^2,g_0)$. All the other $C_i(Q^2)$ quantities are also infinite. To obtain amplitudes in the dressed particle representation consider first a scattering amplitude in the bare particle representation:
\begin{equation}
-i{\cal A}(g_0^2)=g_0^2[D_1(Q^2)+g_0^2D_2(Q^2)+O(g_0^4)].
\end{equation}
We now renormalize $-i{\cal A}(g_0^2)$ in terms of $g^2$:
\begin{equation}
-i{\cal A}(g^2)=g^2[D_1'(Q^2)+g^2D'_2(Q^2)+O(g^4)].
\end{equation}
The invariant scattering amplitude is now determined by the experimental dressed coupling constant $g$. In diagrammatical terms this result is achieved by splitting the $g^4$ term into two pieces, one containing a loop at $Q^2$ and another one at $Q^2=\mu^2$ and the signs of the two terms are opposite. The relation between $g^2$ and $g_0^2$ is determined at a particular value of the virtual boson's momentum transfer, $Q^2=-q^2=\mu^2$, where $\mu$ is the running renormalization group mass parameter.

A derivation of the loop quantity $\Pi(Q^2)$ for $Q^2\le \mu^2$ yields the result
\begin{equation}
\label{Pi2}
\Pi(Q^2)=Q^2\biggl[\frac{\alpha_g}{3\pi}\ln\biggl(\frac{\Lambda_B^2}{Q^2}\biggr)-\frac{\alpha_g}{3\pi}\ln\biggl(\frac{\Lambda_B^2}{\mu^2}\biggr)\biggr]
=Q^2\biggl[\frac{\alpha_g}{3\pi}\ln\biggl(\frac{\mu^2}{Q^2}\biggr)\biggr].
\end{equation}
This is a positive and finite quantity in our dressed particle representation if $\mu^2 > Q^2$. It does not depend on the {\it ad hoc} cutoff $\Lambda_B$ that is taken to the limit $\Lambda_B\rightarrow\infty$. The result shows the difference in $\Pi(Q^2)$ when calculated at scales $Q^2$ and in terms of the running renormalization parameter $\mu^2$ and can be understood in the context of renormalization group flow formalism~\cite{Gellmann}.

Because the scattering amplitude is an observable, it cannot depend on the chosen value of $\mu$. This leads to the renormalization group flow equation:
\begin{equation}
\mu\frac{d{\cal A}}{d\mu}=\biggl(\mu\frac{\partial}{\partial\mu}\vert_g +\mu\frac{\partial g}
{\partial\mu}\frac{\partial}{\partial g}+\mu\frac{\partial g'}
{\partial\mu}\frac{\partial}{\partial g'}\biggr){\cal A}=0.
\end{equation}

The coupling constants are modified in our dressed, renormalized particle representation by the vacuum polarization loops in the propagators. We obtain for the running fine structure constant in QED in the large $Q^2$ limit:
\begin{equation}
\label{alpharunning}
\alpha(Q^2)=\frac{\alpha(\mu^2)}{1-\frac{\alpha(\mu^2)}{3\pi}\ln\biggl(\frac{Q^2}{\mu^2}\biggr)},
\end{equation}
where $\alpha(Q^2)=e^2(Q^2)/4\pi$ and $\mu^2=-q^2=Q^2$ is a particular value of the virtual photon's momentum transfer appropriate for a given experiment. The $\mu^2$ is the physical energy scale at which the running constants are measured.

We obtain from (\ref{Pi2}) for $s/4\sim Q^2$, where $\sqrt{s}$ is the center-of-mass energy, and $4\mu^2 > s$:
\begin{equation}
\label{Pis}
\Pi(s)=s\frac{\alpha_g(s)}{12\pi}\ln\biggl(\frac{4\mu^2}{s}\biggr).
\end{equation}

\section{Symmetry Breaking}

An alternative to the standard perturbative renormalization method is to set the bare boson mass $M_0$ to zero and identify the vector boson self-energy with the boson mass $M^2_V\equiv\delta M_V^2$. We have satisfied Salam's renormalizability condition (\ref{Salamcondition}). The vector boson creates a virtual fermion-anti-fermion pair which in turn creates a vector boson, producing the vector boson self-energy diagram. The fermion-anti-fermion pair can be pictured as a virtual fermion ``condensate''. By incorporating the self-energy contributions, we have shifted from the bare particle representation to the dressed physical representation described by ${\cal L}_0'$ in Eq. (\ref{ShiftedLagrangian}).

Let us now consider a non-Abelian gauge vector field $W_\mu^a$. When written in terms of dressed boson fields, our Lagrangian picks up a quadratic term from the lowest order non-Abelian self-energy diagram:
\begin{equation}
\delta M_W^2=g^2\Pi[T^a\cdot T^b]W^a_\mu W^{\mu b},
\end{equation}
where $\Pi=\Pi(Q^2)$ denotes the proper $W$ boson self-energy contribution. The gauge boson masses squared are determined by the eigenvalues of the 3 by 3 matrix $g^2\Pi[T^a\cdot T^b]$.

Let us consider the symmetry group $G$ which is broken down to the subgroup $H$. We find that $N(G)-N(H)$ Nambu-Goldstone bosons will be generated. We start with $N(G)$ massless gauge bosons, one for each generator. Upon symmetry breaking, the $N(G)-N(H)$ Nambu-Goldstone bosons are eaten by $N(G)-N(H)$ gauge bosons, leaving $N(H)$ massless gauge bosons. For the case of $SU_L(2)\times U_Y(1)$ we have $N(G)=4$ and $N(H)=1$ and we end up with one massless gauge boson, namely, the photon. In our Lagrangian after symmetry breaking:
\begin{equation}
{\cal L}_M=\frac{1}{2}g^2\Pi[T^a\cdot T^b]W^{\mu a}W_\mu^b=\frac{1}{2}W^{\mu a}(M^2)^{ab}W^b_\mu,
\end{equation}
where
\begin{equation}
(M^2)^{ab}=g^2\Pi[T^a\cdot T^b]
\end{equation}
denotes the mass matrix.

We now invoke a dynamical symmetry breaking of $SU(2)\times U(1)$ and diagonalize $(M^2)^{ab}$ to obtain the masses of the gauge bosons. The diagonalized Lagrangian with gauge field interactions is given by
\begin{eqnarray}
\label{WBLagrangian}
{\cal L}_{WB}&=&\Pi\biggl(\frac{1}{2}g^2W^+_\mu W^{-\mu}+\frac{1}{4}g^2W^3_\mu W^{3\mu}-\frac{1}{4}gg'W^3_\mu B^\mu
-\frac{1}{4}gg'B_\mu W^{3\mu} -\frac{1}{4}g^{'2}B_\mu B^\mu\biggr).
\end{eqnarray}
We have
\begin{equation}
\label{AZequations}
B_\mu=A_\mu\cos\theta_w-Z_\mu\sin\theta_w,\quad W^3_\mu=A_\mu\sin\theta_w+Z_\mu\cos\theta_w.
\end{equation}
Substituting (\ref{AZequations}) into (\ref{WBLagrangian}), we obtain
\begin{equation}
\label{WZALagrangian}
{\cal L}_{WZA}=\frac{1}{2}g^2\Pi W^+_\mu W^{-\mu}+\frac{1}{4}g^2\Pi(A_\mu A^\mu\sin^2\theta_w
+A_\mu Z^\mu\sin\theta_w\cos\theta_w
$$ $$
+Z_\mu A^\mu\sin\theta_w\cos\theta_w+Z_\mu Z^\mu\cos^2\theta_w)
-\frac{1}{4}gg'\Pi(A_\mu A^\mu\sin\theta_w\cos\theta_w-A_\mu Z^\mu\sin^2\theta_w
$$ $$
+Z_\mu A^\mu\cos^2\theta_w-Z_\mu Z^\mu\sin\theta_w\cos\theta_w)-\frac{1}{4}gg'\Pi(A_\mu A^\mu\sin\theta_w\cos\theta_w+A_\mu Z^\mu\cos^2\theta_w
$$ $$
-Z_\mu A^\mu\sin^2\theta_w-Z_\mu Z^\mu\sin\theta_w\cos\theta_w)+\frac{1}{4}g^{'2}\Pi(A_\mu A^\mu\cos^2\theta_w-A_\mu Z^\mu\sin\theta_w\cos\theta_w
$$ $$
-Z_\mu A^\mu\sin\theta_w\cos\theta_w+Z_\mu Z^\mu\sin^2\theta_w).
\end{equation}

We are seeking mass terms in the fields, so we ignore mixed terms that describe interactions such as $A_\mu Z^\mu$. We have
\begin{equation}
A_\mu A^\mu\Pi\biggl(\frac{1}{4}g^{'2}\cos^2\theta_w+\frac{1}{4}g^2\sin^2\theta_w-\frac{1}{2}gg'\sin\theta_w\cos\theta_w\biggr).
\end{equation}
By using (\ref{Trigrelations}) we get
\begin{equation}
A_\mu A^\mu\Pi\biggl(\frac{1}{4}g^{'2}\frac{g^2}{(g^2+g^{'2})}+\frac{1}{4}\frac{g^2}{(g^2+g^{'2})}
-\frac{1}{2}gg'\frac{g'}{(g^2+g^{'2})^{1/2}}\frac{g}{(g^2+g^{'2})^{1/2}}\biggr)=0.
\end{equation}
This tells us that the $A_\mu$ photon field is massless. For the $Z$ field we get
\begin{equation}
Z_\mu Z^\mu\Pi\biggl(\frac{1}{4}g^4\frac{g^2}{(g^2+g^{'2})}+\frac{1}{4}g^{'4}\frac{g^2}{(g^2+g^{'2})}+\frac{1}{2}gg'\frac{g'}{(g^2+g^{'2})^{1/2}}
\frac{g}{(g^2+g^{'2})^{1/2}}\biggr)=Z_\mu Z^\mu\Pi\frac{1}{4}(g^2+g^{'2}).
\end{equation}
Therefore, the mass of the $Z$ boson is
\begin{equation}
\label{Zmass}
M_Z=\frac{1}{2}\Pi^{1/2}(g^2+g^{'2})^{1/2}.
\end{equation}
The mass of the $W$ boson is given by
\begin{equation}
\label{Wmass}
M_W=\frac{1}{2}\Pi^{1/2} g.
\end{equation}

From the observed value for Fermi's constant, $G_F=1.166364\times 10^{-5}\, {\rm GeV}^{-2}$, and (\ref{Wmass}) we obtain
\begin{equation}
\Pi(Q^2_{EW})=\frac{4M_W^2}{g^2}=\frac{1}{\sqrt{2}G_f},
\end{equation}
where $Q_{EW}$ is the value of $Q$ corresponding to the EW symmetry breaking energy scale. This yields $\Pi^{1/2}(Q^2_{EW})\sim 246$ GeV for the electroweak scale and with the experimental value, $\sin^2\theta_w=0.2397$, we obtain the boson mass values
\begin{equation}
M_W=\frac{37.3}{\sin\theta_w}\, {\rm GeV}=77.3\, {\rm GeV}, \quad M_Z=\frac{74.6}{\sin2\theta_w}\, {\rm GeV}=87.4\, {\rm GeV}.
\end{equation}

The $\rho$ parameter becomes
\begin{equation}
\rho=\frac{M_W^2}{M_Z^2\cos^2\theta_w}=1.
\end{equation}
We do not identify the proper vector self-energy $\Pi^{1/2}(Q^2_{EW})$ with the standard Higgs model vacuum expectation value $v=\langle\phi\rangle_0$.

\section{Fermion Masses}

 We will generate fermion masses from the one-loop fermion self-energy graph using our self-consistent Hartree-Fock method of approximation~\cite{Nambu}. This method of deriving fermion masses is economical in assumptions, as we obtain the masses from our original massless electroweak Lagrangian by calculating fermion self-energy graphs.

A fermion particle obeys the equation:
\begin{equation}
\label{Dirac}
\slashed p-m_{0f}-\Sigma(p)=0,
\end{equation}
for
\begin{equation}
\label{Dirac2}
\slashed p-m_f=0.
\end{equation}
Here, $m_{0f}$ is the bare fermion mass, $m_f$ is the observed fermion mass and $\Sigma(p)$ is the fermion proper self-energy part. We have
\begin{equation}
m_f-m_{0f}=\Sigma(p,m_f,g,\Lambda_f)\vert_{\slashed p-m_f=0},
\end{equation}
where $\Lambda_f$ denotes the energy scales for individual lepton and quark masses. A solution of (\ref{Dirac}) and (\ref{Dirac2}) can be found by successive approximations starting from the bare mass $m_{0f}$.

The regularized one-loop correction to the self-energy of a fermion with mass $m_f$ and $M_{0V}=0$ is given by~\cite{Evens1991}:
\begin{equation}
-i\Sigma(p)=\int\frac{d^4k}{(2\pi)^4}(ig\gamma_\mu)\frac{-i}{(\slashed{p}-k)-m_f+i\epsilon}(ig\gamma_\nu)\frac{-i\eta^{\mu\nu}}{k^2+i\epsilon}
\exp\left(\frac{(p-k)^2-m_f^2}{\Lambda_f^2}+\frac{k^2}{\Lambda_f^2}\right),
\end{equation}
Promoting the propagator to Schwinger (proper time) integrals using
\begin{equation}
\frac{1}{k^2-m_f^2}=\int_1^\infty\frac{d\tau}{\Lambda_f^2}\exp\left((\tau-1)\frac{k^2-m_f^2}{\Lambda^2}\right),
\end{equation}
and performing the momentum integral, we obtain
\begin{align}
-i\Sigma(p)&=-\frac{ig^2}{8\pi^2}\int_1^\infty d\tau_1\int_1^\infty d\tau_2\left(\frac{\tau_2}{(\tau_1+\tau_2)^3}\slashed{p}+\frac{2}{(\tau_1+\tau_2)^2}m_f\right)
\exp\left(\frac{\tau_1\tau_2}{\tau_1+\tau_2}\frac{p^2}{\Lambda_f^2}-\tau_1\frac{m_f^2}{\Lambda_f^2}\right).
\end{align}
By performing an asymptotic expansion in $\Lambda_f$, we get at $p=0$:
\begin{equation}
\Sigma(0)=m_f\frac{g^2}{4\pi^2}\ln\biggl(\frac{\Lambda_f^2}{m_f^2}\biggr).
\end{equation}
We now identify the fermion mass as $m_f=\Sigma(0)$:
\begin{equation}
m_f=m_f\frac{g^2}{4\pi^2}\ln\biggl(\frac{\Lambda_f^2}{m_f^2}\biggr).
\end{equation}
In addition to admitting a trivial solution at $m_f=0$, this equation also has a non-trivial solution:
\begin{equation}
1=\frac{g^2}{4\pi^2}\ln\biggl(\frac{\Lambda_f^2}{m_f^2}\biggr).
\end{equation}

We use for quarks the strong coupling constant $g_s\simeq 1.5$, and also introduce a color factor 3. We obtain
\begin{equation}
\Lambda_q = 18.5 m_q.
\end{equation}
For a top quark mass $m_t=171.2$ GeV, the corresponding energy scale is $\Lambda_t = 3.2 $ TeV.

It is important to observe that the neutrino is {\it predicted in our EW theory to be massive}, because the neutrino self-energy $\Sigma_\nu$ is non-zero due to the loop calculation composed of an external neutrino with a $W$ or $Z$ loop attached to it. The internal fermion propagator is either associated with an electron or a neutrino depending on whether the boson loop is a $W$ or $Z$ boson, respectively. The massive neutrino has a right-handed partner $\nu_R$ as well as the left-handed partner $\nu_L$. In the standard model with a Higgs mechanism the Higgs coupling to particles is proportional to the mass of the particle. Thus, the standard Higgs particle model can accommodate either a massless neutrino or a massive one. This is not the case in our EW model: the neutrino must have a non-vanishing mass produced by the self-energy $\Sigma_f$. In the standard Higgs model, the massive neutrino with the right-handed $\nu_R$ can present problems with renormalizability due to the need for a 5-dimensional operator incorporating the Higgs field. In contrast, our model with the inclusion of a right-handed neutrino $\nu_R$ can be renormalizable .

In our model, $\Lambda_f$ plays a role that is similar to that of the diagonalized fermion mass matrix in the standard model obtained from a Yukawa Lagrangian with $v=\langle\phi\rangle_0 \neq 0$. The number of undetermined parameters in our EW model is the same as in the standard model: for each fermion, a corresponding $\Lambda_f$ determines its mass. Our model has massive neutrinos. Because the $\Lambda_f$ correspond to the diagonal components of a fermion mass matrix, off-diagonal terms are absent, and no flavor mixing takes place. Therefore, self-energy calculations alone are not sufficient to account for observed neutrino oscillations. However, in addition to fermion self-energy graphs, another case must be considered. Emission or absorption of a charged vector boson $W^\pm$ can be flavor violating through the off-diagonal components of the CKM matrix. In the standard model, such flavor violating terms are not considered significant, due to the smallness of the corresponding CKM matrix elements. But in our model additional factors $\Lambda_{ff'}$ enter into the picture in a manner similar to the self-energy calculation we just described. These may include terms that correspond to the off-diagonal elements of the neutrino mass matrix, offering a natural explanation for neutrino oscillations without having to introduce new interactions.

Further investigations of our model of the origin of quark and lepton masses should be based on self-consistent solutions of the Schwinger-Dyson equations for our fermion Green's functions~\cite{Dyson,Schwinger,Itzykson} and the application of functional renormalization group flow methods~\cite{Gies}.

\section{Tree Graph Unitarity Bound}

Let us consider the Born approximation tree graph calculation of the amplitude for longitudinally polarized $W_L+W_L\rightarrow W_L+W_L$ scattering~\cite{Bell,Smith,Cornwall}. The Feynman tree graphs for the $W_LW_L$ scattering are:
\vskip 16pt\noindent
\begin{equation}
\parbox{1in}{\begin{fmfgraph*}(50,30)
\fmfleftn{i}{2}\fmfrightn{o}{2}
\fmf{boson}{i1,v1}
\fmf{boson}{i2,v1}
\fmf{boson,label=$\gamma/Z^0$,lab.side=left}{v1,v2}
\fmf{boson}{v2,o1}
\fmf{boson}{v2,o2}
\fmflabel{$W^{+}$}{i1}
\fmflabel{$W^{-}$}{i2}
\fmflabel{$W^{+}$}{o1}
\fmflabel{$W^{-}$}{o2}
\end{fmfgraph*}}
+~~~~~~~~~~
\parbox{1in}{\begin{fmfgraph*}(40,40)
\fmfbottomn{i}{2}\fmftopn{o}{2}
\fmf{boson}{i1,v1}
\fmf{boson}{i2,v1}
\fmf{boson,label=$\gamma/Z^0$}{v1,v2}
\fmf{boson}{v2,o1}
\fmf{boson}{v2,o2}
\fmflabel{$W^{+}$}{i1}
\fmflabel{$W^{+}$}{i2}
\fmflabel{$W^{-}$}{o1}
\fmflabel{$W^{-}$}{o2}
\end{fmfgraph*}}\label{eq:WWST}
\end{equation}
\vskip 8pt\noindent
We add to these graphs the $W$ boson contact interaction graph:
\vskip 16pt\noindent
\begin{align}
\parbox{0.55in}{\begin{fmfgraph*}(40,30)
\fmfleftn{i}{2}\fmfrightn{o}{2}
\fmf{boson}{i1,v1}
\fmf{boson}{i2,v1}
\fmf{boson}{v1,o1}
\fmf{boson}{v1,o2}
\fmflabel{$W^-_\lambda$}{i1}
\fmflabel{$W^+_\mu$}{i2}
\fmflabel{$W^-_\rho$}{o1}
\fmflabel{$W^+_\nu$}{o2}
\end{fmfgraph*}}\label{eq:4W}\\
&\nonumber
\end{align}
\vskip 6pt\noindent

The amplitude for the longitudinally polarized $W$ scattering is given by~\cite{MoffToth}:
\begin{equation}
\label{WWamplitude}
{\cal A}(W_LW_L\rightarrow W_LW_L)=g^2\left[\frac{\cos\theta+1}{8M_W^2}s+{\cal O}(1)\right],
\end{equation}
where $g$ is the low energy weak coupling constant and $\theta$ is the scattering angle. Eq. (\ref{WWamplitude}) clearly violates unitarity for large $s$. However, this behavior is corrected in the standard Weinberg-Salam model by the addition of the Higgs exchange process:
\vskip 16pt\noindent
\begin{align}
\parbox{0.55in}{\begin{fmfgraph*}(50,30)
\fmfleftn{i}{2}\fmfrightn{o}{2}
\fmf{boson}{i1,v1}
\fmf{boson}{i2,v1}
\fmf{dashes,label=$H$}{v1,v2}
\fmf{boson}{v2,o1}
\fmf{boson}{v2,o2}
\fmflabel{$W^-_\lambda$}{i1}
\fmflabel{$W^+_\mu$}{i2}
\fmflabel{$W^-_\rho$}{o1}
\fmflabel{$W^+_\nu$}{o2}
\end{fmfgraph*}}\label{eq:WWH}\\
&\nonumber
\end{align}
\vskip 8pt\noindent
In the high energy limit we get
\begin{equation}
\label{Higgscontribution}
{\cal A}_H=-g^2\left[\frac{\cos\theta+1}{8M_W^2}s+{\cal O}(1)\right],
\end{equation}
which cancels the bad high energy behavior in (\ref{WWamplitude}). The resulting scattering amplitude in the standard EW model based on EW spontaneous symmetry breaking, including the Higgs particle exchange graph, is given by
\begin{equation}
\label{WWstandardmodel}
{\cal A}_\mathrm{SM}(W_LW_L\rightarrow W_LW_L)=g^2\left[\frac{\cos^2\theta+3}{4\cos\theta_w^2(1-\cos\theta)}-\frac{M_H^2}{2M_W^2}+{\cal O}(s^{-1})\right].
\end{equation}
We observe that if we integrate over the scattering angle $\theta$ in (\ref{WWstandardmodel}) to obtain the cross section, then there is no energy dependence to order ${\cal O}(s^{-1})$. The fundamental Higgs mass cannot be bigger than $M_H\sim 800$ GeV to prevent a break down of perturbation theory and unitarity.

Our EW theory does not have a physical Higgs particle, so we must consider an alternative way of avoiding unitarity violation. Substituting the running $W$ mass
\begin{equation}
M^2_W(s)=\frac{1}{4}g^2\Pi(s)
\end{equation}
into (\ref{WWamplitude}), we find that
\begin{equation}
\label{WWamplitude2}
{\cal A}(W_LW_L\rightarrow W_LW_L)=\left[\frac{\cos\theta+1}{2\Pi(s)}s+{\cal O}(1)\right].
\end{equation}
The running coupling constant $\alpha_g(s)$ has the behavior for large $4\mu^2 > s$:
\begin{equation}
\label{alphag}
\alpha_g(s)=\frac{\alpha(\mu^2)}{1-\frac{\alpha(\mu^2)}{3\pi}\ln\biggl(s/4\mu^2\biggr)}\sim \frac{3\pi}{\ln\biggl(4\mu^2/s\biggr)}.
\end{equation}
Moreover, $\Pi(s)$ has the behavior for large $4\mu^2 > s$:
\begin{equation}
\label{Pis2}
\Pi(s)\sim s\frac{\alpha_g(s)}{12\pi}\ln\biggl(\frac{4\mu^2}{s}\biggr).
\end{equation}
Now it follows that 
\begin{equation}
\Pi(s)\sim s/4\pi
\end{equation}
and the amplitude ${\cal A}(W_LW_L\rightarrow W_LW_L)$ is bounded by a constant as $s\rightarrow\infty$ and unitarity is not violated. The scattering amplitude ${\cal A}(W_LW_L\rightarrow W_LW_L)$ has been calculated in the dressed tree graph representation in which the running of the coupling constant and the mass of the $W$ particle are accounted for.

\end{fmffile}

\section{Conclusions}

The conditions for renormalizability of general gauge theories of massive vector bosons derived by Salam~\cite{Salam2} show that, with the exception of neutral vector bosons, non-Abelian gauge bosons are non-renormalizable unless the bare boson mass $M_{0V}=0$. The physical boson masses can be calculated by using a self-consistent Hartree-Fock method. Besides the trivial solution with $M_{0V}=0$, there is a second solution for $M_V\neq 0$ based on expanding the fields around states with ``dressed'' physical particles described by the Lagrangian: ${\cal L}_0'={\cal L}_0+{\cal L}_{\rm self}$. This Lagrangian is associated with different physical states and a different vacuum state, replacing those with ``bare'' particle states described by the Lagrangian ${\cal L}_0$. The vector boson mass is identified with the proper self-energy $M_V^2=\delta M^2_V$. In the bare particle basis the EW symmetry remains unbroken. Physical particles must have mass and thus cannot have the gauge symmetry of the underlying bare particle representation. The symmetry of the EW group $SU_L(2)\times U_Y(1)$ is dynamically broken to $U_{\rm EM}(1)$, retaining a massless photon and generating the masses of the $W$ and $Z$ bosons. Nambu-Goldstone bosons become the scalar longitudinal components of the three transverse degrees of freedom of the massive $W$ and $Z$ vector bosons. The theory is renormalizable and there exist Ward-Takahashi identities~\cite{Aitchison} corresponding to a conserved current.

Physical fermion masses are generated from a Nambu-Jona-Lasinio~\cite{Nambu} self-consistent scheme. The fermion bare masses $m_{0f}=0$ and the physical masses are computed from the proper self-energies $m_f\equiv\Sigma_f=\delta m_f$. Whereas in the standard Weinberg-Salam model the physical masses are generated in the tree graph, Born approximation by introducing a classical scalar field $\phi$ and invoking a spontaneous symmetry breaking through a non-vanishing vacuum expectation value $v=\langle\phi\rangle_0\neq 0$, in our EW theory the masses of the elementary fermions and bosons are generated in a self-consistent dynamical scheme from the purely quantum loop calculations of the self-energies of the fields. The tree graph, Born approximation unitarity bound for the amplitude for polarized longitudinal $W_LW_L$ scattering is not violated due to the running of the $W$ boson mass $M_W(s)\sim s$ as $s\rightarrow\infty$. The tree graph approximation is calculated in our physical particle representation with the fully dressed particles and the boson masses generated by the particle self-energies.

A further study must be conducted to determine the exact renormalization group flow equations and to calculate the fixed points in the $\beta$ functions produced by our EW theory. Our EW theory is based on a minimal particle scheme consisting of the observed 12 quarks and leptons, the $W$ and $Z$ bosons and the massless photon and gluons. There is no fundamental Higgs particle in the physical particle spectrum, our EW theory does not have a Higgs hierarchy mass problem and there is no severe fine tuning associated with the Higgs mechanism vacuum energy density.

\section*{Acknowledgements}

I thank John Dixon, Martin Green and Viktor Toth for helpful and stimulating discussions. This research was generously supported by the John Templeton Foundation. Research at the Perimeter Institute for Theoretical Physics is supported by the Government of Canada through NSERC and by the Province of Ontario through the Ministry of Research and Innovation (MRI).

\end{document}